\documentclass[superscriptaddress,showpacs,twocolumn,aps,prl]{revtex4-1}

\usepackage{graphicx,dcolumn,bm,amsmath,color,amssymb}
\usepackage{units}

\newcommand{\bhu}{ \hat{\bf u} }

\newcommand{\br}{ {\bf r} }

\newcommand{\ga}{ {\alpha }}
\newcommand{\gb}{ {\beta }}

\newcommand{\ChangesKaiser}{ \textcolor{red}}

\begin{document}

\title{Transport powered by bacterial turbulence}

\author{Andreas Kaiser}
\email{kaiser@thphy.uni-duesseldorf.de}
\affiliation{Institut f\"ur Theoretische Physik II: Weiche Materie,	Heinrich-Heine-Universit\"at D\"{u}sseldorf, 
D-40225 D\"{u}sseldorf, Germany}

\author{Anton Peshkov}
\affiliation{Laboratoire de Physique et M\'ecanique des Milieux H\'et\'erog\'enes, Ecole Sup\'erieure de Physique et de Chimie Industrielles de la Ville de Paris, 75231 Paris Cedex 05,  France
}
\affiliation{Materials Science Division, Argonne National Laboratory, Illinois 60439, USA}

\author{Andrey Sokolov}
\affiliation{Materials Science Division, Argonne National Laboratory, Illinois 60439, USA}

\author{Borge ten Hagen}
\affiliation{Institut f\"ur Theoretische Physik II: Weiche Materie,	Heinrich-Heine-Universit\"at D\"{u}sseldorf, 
D-40225 D\"{u}sseldorf, Germany}

\author{Hartmut L\"owen}
\affiliation{Institut f\"ur Theoretische Physik II: Weiche Materie,	Heinrich-Heine-Universit\"at D\"{u}sseldorf, 
D-40225 D\"{u}sseldorf, Germany}

\author{Igor S. Aranson}
\email{aronson@anl.gov}
\affiliation{Materials Science Division, Argonne National Laboratory, Illinois 60439, USA}

\date{\today}

\begin{abstract}
We demonstrate that collective turbulent-like motion in a bacterial bath can 
power and steer directed transport of mesoscopic
carriers through the suspension. In our experiments and simulations,
a microwedge-like ``bulldozer''  draws energy from  a bacterial
bath of varied density. We obtain that a maximal transport speed
is achieved  in the turbulent state of the bacterial suspension.
This apparent rectification of random motion of bacteria 
is caused by polar ordered  
bacteria inside the cusp region of the carrier,  
which is shielded from the outside turbulent fluctuations. 
\end{abstract}


\maketitle


{\it Introduction.} Suspensions of bacteria or synthetic microswimmers show fascinating
collective behavior emerging from their self-propulsion
~\cite{Marchetti_Rev,Cates_Rev2012,Romanczuk2012,aranson_ufn}
which results in many novel active states such as swarming
~\cite{Ramaswamy_Science,Gompper,2012SwinEtAl,Ginelli_PRL10}
and ``active turbulence''
~\cite{Sokolov_PRL12,2008Saint_Shelley,Wensink_PNAS,Li_PRE,yang2014,zhou2014}.
In contrast to hydrodynamic turbulence,
the apparent turbulent (or swirling) state occurs at exceedingly low Reynolds numbers but
at relatively large bacterial concentrations.

Here we address the question whether one can systematically extract energy
out of the  seemingly turbulent state established by swimming bacteria and how  the bacterial turbulence
may power micro-engines and transport mesoscopic
carriers  through the suspension. A related question is what processes on a scale of an individual 
swimmer are responsible for the energy rectification from this ``active heat bath''. 

In our experiments we analyze the motion of a microwedge-like carrier (``bulldozer'') 
submersed in a suspension of swimming bacteria  \emph{Bacillus subtilis}. 
Experimental studies are combined  with  particle-resolved computer simulations.
A broad span of bacterial densities is examined,
ranging from the dilute regime over the turbulent to the jammed state.
Due to the activity of the suspension,
the bulldozer-like particle is set  into a rectified motion
along its wedge cusp
~\cite{chaikin,ReichhardtPRL}, in contrast to tracers with symmetric shape~\cite{Clement_PRL11}.
Its averaged propagation speed becomes maximal within the turbulent-like  regime of collective swimming.
Our simulations and experiment indicate that the directed motion  is caused by 
polar ordered bacteria trapped inside the carrier in a region near the cusp
which is shielded from outer turbulent fluctuations. 
The orientation of  trapped bacteria 
yields a double-peaked distribution centered in the direction of the average carrier motion.
Consequently, the  bacterial turbulence powers efficiently the
transport of carriers through the suspension.
This finding opens the way to utilize self-propulsion
energy of bacteria forming a turbulent active fluid for the purpose of 
 control and powering  of mesoscopic engines.

Converting bacterial self-propulsion into mechanical energy  has been
considered previously for shuttles and cogwheels 
~\cite{DiLeonardoPRL,AngelaniCARGO,SokolovPNAS,Wong,wensink2014,di2010bacterial}. 
Most of the studies were restricted to low swimmer concentrations where swirling is absent. 
While fundamental microscopic mechanisms of energy transduction and interaction with solid walls 
on a scale of a single bacterium are fairly well understood, see 
e.g.~\cite{lauga2006swimming,li2009accumulation, drescher2011fluid}, the role of collective motion 
has not been elucidated so far. Ref.~\cite{ishikawa2011energy} demonstrated that  while the energy 
generated by individual bacteria dissipates on microscale,  the increase in swimming velocity and mass 
transport due to collective motion is significant. 
Here we put forward an idea
how the collective  bacterial swimming (bacterial turbulence)  strongly amplifies the energy
transduction.  Complementary, in high-Reynolds-number turbulent flows, the motion of suspended
inertial particles is not directed \cite{Lohse}, lacking a conversion
of turbulent fluctuations into useful mechanical work.

\begin{figure}[tb]
\begin{center}
\includegraphics[clip=,width=1\columnwidth]{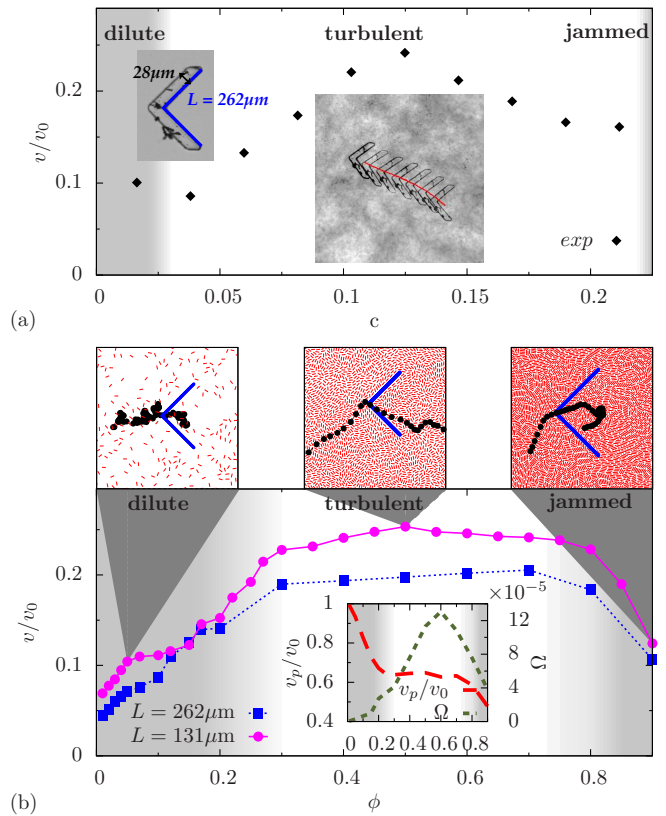}
\caption{ \label{f1}
(a) Experimental carrier speed $v/v_0$ as a function of the bacterial 3D volume fraction   $c$. 
The insets show the temporal progress of the carrier 
positions (right) as well as a snapshot of the carrier (left) indicating the characteristic 
spatial dimensions and the schematic representation (blue line). See also Supplementary Movie 1. 
(b) Numerically obtained transport speed for varying swimmer concentration  $\phi$ and carriers of two different contour lengths:  
 $L=\unit[262]{\mu m}$ (squares) corresponding to the experimental size of the carrier and half the length
$L=\unit[131]{\mu m}$ (circles). Magnitude of vorticity $\Omega$ and averaged bacterial swimming speed $v_{p}/v_0$  for various concentrations are
shown in the inset. See also Supplementary Movies  2 and 3.}
\end{center}
\end{figure}

{\it Experiment.} Mesoscopic wedge-like carriers were fabricated by photolithography
\cite{SokolovPNAS,Clement_PRL13}. 
We mixed a liquid photoresist SU-8 with micron-size magnetic particles before spin coating.
This allows to control the orientation of the carriers 
with an external magnetic field applied parallel to the fluid surface.
The  arm length of the  wedge-like carriers is $L= \unit[262]{\mu m}$ (see Fig.~\ref{f1}),
and the wedge angle is 90$^{\circ}$.
Experiments were conducted on a suspension of \emph{Bacillus subtilis}, a flagellated rod-shaped
swimming bacterium $\ell \sim \unit[5]{\mu m}$ long and  $\unit[0.7]{\mu m}$ wide. The suspension
of bacteria was grown for 8--12 hours in Terrific Broth growth medium (Sigma Aldrich). To monitor the
concentration of bacteria during the growth phase, we continuously measured the optical scattering of
the medium using an infrared proximity sensor. At the end of the exponential growth phase the bacteria were washed
and centrifuged to achieve the desired concentration. Then a small drop of concentrated bacterial
suspension was placed between four movable fibers and stretched up to the thickness of $\sim \unit[100]{\mu m}$
(see Ref.~\cite{Sokolov_PRL07}). Both surfaces of the free-standing  liquid film were exposed to air,
significantly increasing the oxygen diffusion rate into the bulk of the film. According to our previous
study~\cite{Sokolov_PRL12}, a relatively high concentration of oxygen is required for bacterial motility
in concentrated suspensions of \emph{Bacillus subtilis}. We measured the bacteria concentration (or, equivalently, three-dimensional volume fraction) by means of 
optical coherence tomography (see Ref.~\cite{Sokolov_PRE09}) before and after the experiments in order  
to monitor the effect of film evaporation. 
This revealed that evaporation is negligible in the course of our relatively short experiment.

Two pairs of orthogonal Helmholtz coils were used to create
 a uniform magnetic field in the bulk of the liquid film. The
carrier was carefully inserted in the  film by  a digital micropipet. In the  course of our experiment, the
orientation of the magnetic carrier was reversed every $\unit[20]{s}$ to prevent migration of  the carrier out
of the field of view. We also confirm that the average speed of the wedge does not depend on the direction of the 
motion. The influence of gravity  was negligible in our experiment. The motion of the wedge 
was captured by a digital high-resolution microscope camera 
[Fig.~\ref{f1}(a) and the Supplemental Material~\cite{SuppMat}] for the duration of $\unit[2-4]{min}$. Both 
displacement and orientation of the carrier were tracked by a custom-designed
software based on Matlab toolboxes.

{\it Simulation.} We model the bacteria by rod-like objects with repulsive interactions
and an effective self-propulsion using parameters matching the experimental conditions.
More specifically, the excluded volume interaction between the rods is described by
 $n$  ``Yukawa'' segments positioned equidistantly with the distance $d=\unit[0.85]{\mu m}$ along a stiff axis of length $\ell$,
i.e., a repulsive Yukawa potential is imposed between the segments of different rods~\cite{Kirchhoff1996}.
However, in order to properly take into account collisions between the bacteria, an important modification is introduced 
in the model compared to that of Ref.~\cite{Wensink_PNAS}. 
Experiments~\cite{Sokolov_PRL07,aranson2007,drescher2011fluid} demonstrated that two 
bacteria swim away from each other after the collision. This effect results in a 
suppression of clustering for small and moderate bacteria concentrations. However, in previous 
simulations~\cite{Wensink_PNAS}, 
the bacteria had a propensity to swim parallel after the collision and 
to form dense clusters with a smectic-like alignment.
In order to describe the experimentally observed swim-off effect and the resulting suppression of clustering, 
we incorporate an asymmetric effective bacterial shape by enlarging the interaction prefactor of 
the first segments of each rod with respect to the other segments by a factor of 3 (see Supplemental Material~\cite{SuppMat}).
The resulting total interaction potential between a swimmer pair ${\ga,\gb}$ is then given
by $U_{\ga \gb} = \sum_{i=1}^{n}\sum_{j=1}^{n}{U_i U_j} \exp [-r_{ij}^{\ga \gb} / \lambda]/r_{ij}^{\ga \gb}$
with $U_1^2 / U_j^2 = 3$ ($j=2  \ldots n$),
where $\lambda$ is a screening length  obtained from the  experimental
effective rod aspect ratio $\ell/\lambda=5$, and
$r_{ij}^{\alpha \beta} = |{\bf r}_{i}^{\ga} - {\bf r}_{j}^{\gb}|$ is the distance between segment
$i$ of rod $\alpha$ and segment $j$ of rod $\beta$ ($\alpha \neq \beta $). 
The carrier is implemented correspondingly by tiling the wedge-like contour with length
$L=\unit[262]{\mu m}$ [see Fig.~\ref{f1}(a)] with  Yukawa segments. The ratio $L/\ell$ is matched to the experimental
situation. For comparison we also perform simulations for smaller arm lengths.
The self-propulsion is taken into account via a formal effective force $F_{0}$ acting
along the rod axis $\bhu = (\cos \varphi,\ \sin \varphi)$.
By imposing a large interaction strength $U_j^{2}=2.5F_{0}\ell$,
we ensure that the bacteria and the
wedge do not overlap. The model neglects long-range hydrodynamic interactions between the swimmers. 
These  interactions do not change the overall morphology of the bacterial flow. 
The most noticeable effect of the hydrodynamic interactions is a seven to ten fold increase of the 
collective flow velocity compared to the speed of individual bacteria \cite{dombrowski2004self,Sokolov_PRL07}.  
Certainly, this phenomenon cannot be attained by our model and would require proper incorporation of the hydrodynamic 
forces.

Since bacterial swimming occurs  at  exceedingly low Reynolds numbers, 
the overdamped  equations of motion for the positions
and orientations of the rods are
\begin{eqnarray}
{\bf f }_{\cal T} \cdot \partial_{t} \br_{\alpha}(t) &=&  -\nabla_{\br_{\alpha}}
 U +  F_{0} \bhu_{\alpha}(t)\,, \\
{\bf f}_{\cal{R}} \cdot \partial_{t} \bhu_{\alpha}(t) &=&
-\nabla_{\bhu_{\alpha}} U \,.
\label{eom:rods}
\end{eqnarray}
Here, $U=(1/2)\sum_{\alpha, \beta (\alpha \neq
  \beta)} U_{\alpha \beta} + \sum_{\alpha} U_{\alpha <}$ is the total potential energy,
where $U_{\alpha <}$  denotes the interaction  energy of rod $\alpha$ with the
carrier. (In general, a subscript $<$ refers to a
quantity associated with the wedge-like carrier.) 
The one-body translational and rotational friction tensors for the rods ${\bf f}_{\cal T}$
and ${\bf f}_{\cal R}$ can be decomposed into parallel $f_{\parallel}$, perpendicular $f_\perp$,
and rotational $f_R$ components which depend solely on the
aspect ratio $\ell/\lambda $ and are taken from Ref.~\cite{tirado}.
The resulting self-propulsion speed of a single rod  $v_{0}=F_{0}/f_{\parallel}$
 is matched to the experimental value $\unit[15]{\mu m / s}$~\cite{Sokolov_PRL07}
 leading to the time unit $\tau = \ell / v_0$.
Since at a relatively large bacteria concentration
thermal fluctuations and tumbling  are not important, 
we neglect all stochastic noise terms (our experimental studies in Ref.~\cite{Sokolov_PRL12} showed that tumbling of 
{\it Bacillus subtilis} becomes significant only for very low oxygen concentrations). 
Moreover, details of the hydrodynamics between the bacteria 
and the air-water interface are neglected.
According to the experiment, the motion of the carrier is mostly  translational 
and induced by the carrier-bacteria interactions.
The hydrodynamic friction tensor ${\bf f }_{<}$ of the wedge is calculated for the specific
geometry with the dimensions shown in the sketch (left inset) in Fig.~\ref{f1}(a). For this purpose, the shape of
the carrier is approximated by a large number of beads that are rigidly connected. The corresponding hydrodynamic
calculations based on the Stokes equation for the flow field around a  particle at low Reynolds
number are  performed with the software package \texttt{HYDRO++} \cite{delaTorreNMDC1994,Carrasco99}.

The resulting equation of motion for the carrier is
\begin{eqnarray}
{\bf f }_{<} \cdot \partial_{t} \br_{<}(t) &=&  -\nabla_{\br_{<}} \sum_{\alpha}{U_{\alpha <}(t)}\,.
\label{eom:carrier}
\end{eqnarray}
We simulate $N \sim 10^4$ rods  and a single carrier in a square simulation domain with the area 
$ A= ( 3L / \sqrt{2})^{2}$ and periodic boundary conditions in both directions. 
The dimensionless packing fraction $\phi = N \lambda \ell / A$ corresponds to the bacterial volume fraction  $c$ in the experiments.

{\it Results and Discussion.} The shape reflection symmetry of the  wedge around its apex will exclude any
averaged directed motion perpendicular to the apex
while there is no such symmetry in the apex direction. Hence, due to  rectification of random fluctuations,
the carrier will proceed on average
along its cusp. The transport efficiency of the carrier can then be characterized
by its average migration speed $v$ in this direction.
We have examined the carrier motion in a wide range of bacterial bulk concentrations
 including a dilute regime, where bacterial swimming is almost uncorrelated, as well as an intermediate
 turbulent and a final jammed regime.
These regimes can be characterized by suitable order parameters.
For that purpose, we define the mean magnitude of vorticity
$\Omega = \frac{1}{2} \langle | [\nabla \times \textbf{V}(\br,t)] \cdot \hat{{\bf e}}_{z}| ^{2} \rangle$ 
for a bacterial velocity field $\textbf{V}(\br,t)$ coarse-grained over three bacterial lengths 
which is a convenient  indicator for
bacterial turbulence~\cite{Wensink_PNAS,WensinkJPCM}. The average swimming speed $v_{p}$ of the bacteria
obtained by averaging the displacements after a time $t=10^{-3}\tau$
indicates jamming at high concentrations.
Simulation results for these two order parameters are presented for a  bacterial suspension in the
absence of the carrier [see inset in Fig.~\ref{f1}(b)]. The results indicate three different states: 
``dilute'' ($\phi \lesssim 0.25$), ``turbulent'' ($0.25 \lesssim \phi \lesssim 0.75$), and ``jammed''
($0.75 \lesssim \phi$).  The same sequence of states is found in the experiments~\cite{Sokolov_PRL09}.

Figures~\ref{f1}(a) and~\ref{f1}(b) show that the transport efficiency $v/v_0$ of the carrier
peaks in the turbulent regime where it attains a significant fraction
of the net bacterial velocity $v_{0}$. Experimentally, this fraction is found to be
about 0.25,  which is confirmed by the simulations.
Snapshots from the experiments and simulations
(see insets in Fig.~\ref{f1} and the Supplemental Material~\cite{SuppMat}) show a directed motion along the wedge
apex though there are considerable fluctuations which we discuss later.
For a very  dilute regime, there are only a few bacteria pushing the
carrier such that $v$ tends to zero in this limit 
[note that in the experiment  no motion of the carrier was observed in a  very 
dilute regime since the resulting bacterial forces are 
 not sufficient  to overcome the friction  of the carrier with the surface (air-water interface)].
We have also performed simulations for different carrier lengths and  opening angles
to determine the geometry leading to optimal transport.
While we have chosen the optimal apex angle for our experiments, a slightly higher transport 
efficiency can be achieved with a smaller carrier length, see Supplemental Material~\cite{SuppMat}.

\begin{figure}[tb]
\begin{center}
\includegraphics[clip=,width=1\columnwidth]{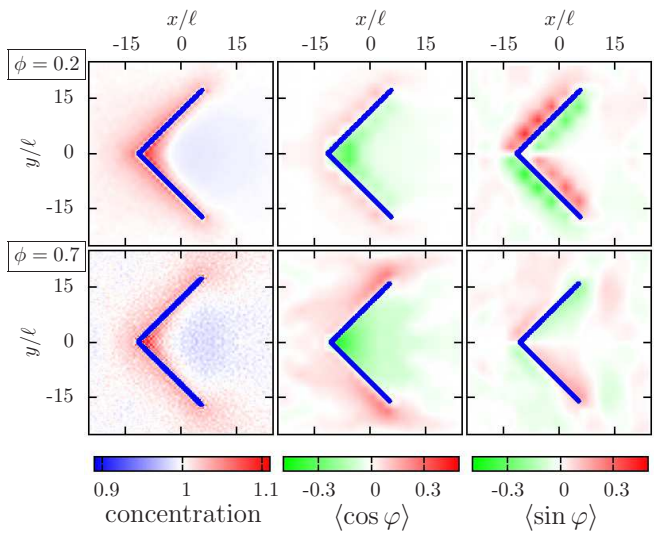}
\caption{\label{contour} Intensity plots for different bacteria concentrations, $\phi=0.2$ (top row) and $\phi = 0.7$ (bottom row):
local bacterial concentration around the carrier, normalized by the total concentration (left) as well as 
averaged bacteria orientations $\langle \cos \varphi \rangle$ (middle) and $\langle \sin  \varphi \rangle$ (right).}
\end{center}
\end{figure}

\begin{figure}[tb]
\begin{center}
\includegraphics[clip=,width=1\columnwidth]{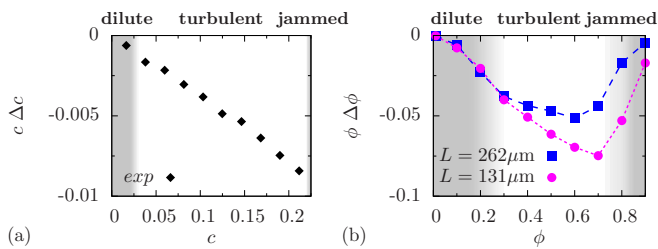}
\caption{\label{FigDensity} Concentration difference between the wake of the carrier and its front $\Delta c = c_w - c_f$ obtained from (a) experiments and (b) simulations.}
\end{center}
\end{figure}

\begin{figure}[tb]
\begin{center}
\includegraphics[clip=,width= 0.86\columnwidth ]{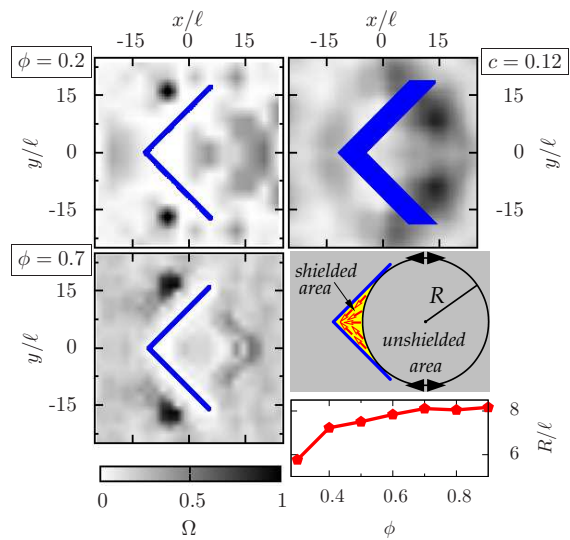}
\caption{ \label{MagnVort} Normalized local magnitude of vorticity obtained from simulations 
(left column) and experiment (top right) for
given bacterial concentrations. Bottom right: illustration of swirl shielding 
in the carrier cusp---bacteria in the shielded area (light colored) are indicated by arrows
and the unshielded area is marked by dark color---as well as typical swirl radii $R$ for different 
bacteria concentrations in the turbulent regime,
obtained as the  first minimum of the equal-time spatial velocity autocorrelation function~\cite{WensinkJPCM}.}
\end{center}
\end{figure}

In the following we discuss the underlying reason for the optimal carrier transport in the turbulent regime.
First, the bacteria inside the wedge close to the cusp are on average orientationally ordered along
the wedge orientation ($x$-direction). The orientational ordering  is revealed by the intensity plots 
for $\langle \cos \varphi \rangle$, $\langle \sin \varphi \rangle$which is used  as
an appropriate orientational order parameter (see the green ``hot spot'' in Fig.~\ref{contour}). The ``hot spot'' sets the
carrier into motion along the $x$-direction.
Similar to a  moving bulldozer piling up sand, the carrier motion  causes an accumulation 
of  particles in the front 
and a depleted wake,  while not destroying the ordering of particles inside the wedge (see the intensity plot in Fig.~\ref{contour} as well as Fig.~\ref{FigDensity} for experimental and simulation data for the
averaged concentration difference between inside and outside bacteria). This concentration  difference decreases
the transport speed. But the driving effect increases  with the increase in  bacterial concentration.
When the turbulence sets in,  there is  a shielding of turbulent fluctuations  near  the 
walls of the carrier (see the intensity plots of the local magnitude of vorticity in Fig.~\ref{MagnVort}).
The shielding is, however, more pronounced inside than outside the wedge.
Intuitively this implies that the hot spot is shielded from swirls which would sweep away 
the driving bacteria. (Concomitantly, the outside swirls shown in Fig.~\ref{MagnVort} are induced by the bulldozer motion
but do not cause the motion.) The intuitive concept  of swirl shielding is sketched in Fig.~\ref{MagnVort}
where also a typical swirl size as a function of the density is shown. A typical swirl of this size
can never reach the ``hot spot" area as
schematically shown  by the shielded area in Fig.~\ref{MagnVort}. 
Swimmer within this area are trapped, leading to large transport velocity correlation
times for the carrier, see Ref.~\cite{SuppMat}.
For simple geometric reasons, there is  no such swirl
shielded zone for the outside bacteria as  a swirl can sweep them away.
At very high concentrations $(\phi \approx 0.8)$,
the bacteria are jammed (see Fig.~\ref{f1}), which is manifested  also in a reduced carrier mobility.
In conclusion, the polar order of bacteria (see intensity plots in Fig.~\ref{contour}) 
inside the wedge and its shielding
from the swirls  are the two basic ingredients to understand the optimal transport
in the turbulent state.
As a consequence, optimal transport is achieved when the carrier aperture width  is  comparable  to
 typical swirl size, see the Supplemental Material~\cite{SuppMat}.

\begin{figure}[tb]
\begin{center}
\includegraphics[clip=,width=1\columnwidth]{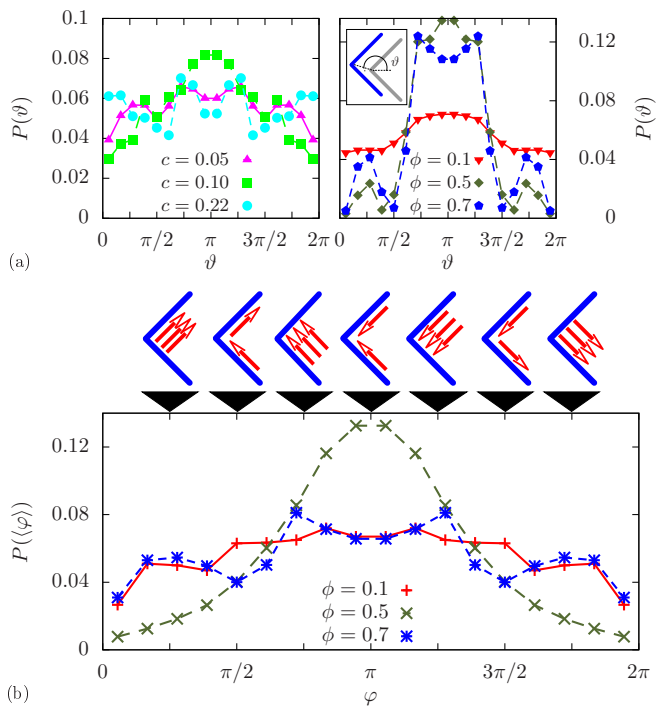}
\caption{\label{FigAngles} (a) Symmetric full probability distributions of the carrier displacement direction $\vartheta$:
Experimental results (left) measured after a time of $\unit[0.03]{s}$ and numerically obtained data (right) 
after a time of $10^{-3}\tau$ for various
bacterial concentrations. The definition of the displacement angle $\vartheta$ is  illustrated  in the inset.
(b) Orientational order distribution of the bacteria within the swirl shielded area.
Characteristic configurations are illustrated by sketches.}
\end{center}
\end{figure}

We  plot the full distribution of the carrier velocity direction $P (\vartheta)$ in Fig.~\ref{FigAngles}(a).
For small bacterial densities $\phi$ the distribution is random, while for intermediate concentrations this 
distribution exhibits a single peak centered around the $x$-axis.
For even higher concentrations it becomes double-peaked corresponding to a motion perpendicular to the
single wedge walls $( \pi \pm \pi/4)$.
This is correlated with the orientational order distribution of the
inside bacteria in the shielded area, $P(\langle \varphi \rangle)$, implying that there is a flipping in the orientation
of the inside bacteria, see Fig.~\ref{FigAngles}(b), for high bacterial concentrations. 
The kink-like change in the direction of motion perpendicular to the wedge walls 
is also observed experimentally [see the trajectory in Fig.~\ref{f1} and peaks in Fig.~\ref{FigAngles}(a)].

{\it Conclusion.} 
We have shown that mechanical energy of bacterial turbulent movements can be extracted to power
directed motion of  a wedge-like carrier.
Both polar ordering  and swirl shielding inside the wedge yield an optimal transport velocity
which becomes even bidirectional at high concentrations.
This effect can be exploited to power and steer carriers and motors
by bacterial turbulence or collective motion of synthetic swimmers.

\acknowledgments 
Work by AK and BtH  was supported by the ERC Advanced Grant INTERCOCOS (Grant No. 267499) and HL was supported by the SPP 1726 of the DFG.
Work by  AS, AP, and ISA was supported by the US Department of Energy (DOE), Office of Science, Basic Energy Sciences (BES), 
Materials Science and Engineering Division.

\bibliographystyle{apsrev}
\bibliography{refs}

\begin{thebibliography}{40}
\expandafter\ifx\csname natexlab\endcsname\relax\def\natexlab#1{#1}\fi
\expandafter\ifx\csname bibnamefont\endcsname\relax
  \def\bibnamefont#1{#1}\fi
\expandafter\ifx\csname bibfnamefont\endcsname\relax
  \def\bibfnamefont#1{#1}\fi
\expandafter\ifx\csname citenamefont\endcsname\relax
  \def\citenamefont#1{#1}\fi
\expandafter\ifx\csname url\endcsname\relax
  \def\url#1{\texttt{#1}}\fi
\expandafter\ifx\csname urlprefix\endcsname\relax\def\urlprefix{URL }\fi
\providecommand{\bibinfo}[2]{#2}
\providecommand{\eprint}[2][]{\url{#2}}

\bibitem[{\citenamefont{Marchetti~et al}(2013)}]{Marchetti_Rev}
\bibinfo{author}{\bibfnamefont{M.~C.} \bibnamefont{Marchetti~et al}},
  \bibinfo{journal}{Rev. Mod. Phys.} \textbf{\bibinfo{volume}{85}},
  \bibinfo{pages}{1143} (\bibinfo{year}{2013}).

\bibitem[{\citenamefont{Cates}(2012)}]{Cates_Rev2012}
\bibinfo{author}{\bibfnamefont{M.~E.} \bibnamefont{Cates}},
  \bibinfo{journal}{Rep. Prog. Phys.} \textbf{\bibinfo{volume}{75}},
  \bibinfo{pages}{042601} (\bibinfo{year}{2012}).

\bibitem[{\citenamefont{Romanczuk et~al.}(2012)\citenamefont{Romanczuk, B\"ar,
  Ebeling, Linder, and Schimansky-Geier}}]{Romanczuk2012}
\bibinfo{author}{\bibfnamefont{P.}~\bibnamefont{Romanczuk}},
  \bibinfo{author}{\bibfnamefont{M.}~\bibnamefont{B\"ar}},
  \bibinfo{author}{\bibfnamefont{W.}~\bibnamefont{Ebeling}},
  \bibinfo{author}{\bibfnamefont{B.}~\bibnamefont{Linder}}, \bibnamefont{and}
  \bibinfo{author}{\bibfnamefont{L.}~\bibnamefont{Schimansky-Geier}},
  \bibinfo{journal}{Eur. Phys. Lett. Spec. Top.}
  \textbf{\bibinfo{volume}{202}}, \bibinfo{pages}{1} (\bibinfo{year}{2012}).

\bibitem[{\citenamefont{Aranson}(2013)}]{aranson_ufn}
\bibinfo{author}{\bibfnamefont{I.~S.} \bibnamefont{Aranson}},
  \bibinfo{journal}{Physics-Uspekhi} \textbf{\bibinfo{volume}{56}},
  \bibinfo{pages}{79} (\bibinfo{year}{2013}).

\bibitem[{\citenamefont{Narayan et~al.}(2007)\citenamefont{Narayan, Ramaswamy,
  and Menon}}]{Ramaswamy_Science}
\bibinfo{author}{\bibfnamefont{V.}~\bibnamefont{Narayan}},
  \bibinfo{author}{\bibfnamefont{S.}~\bibnamefont{Ramaswamy}},
  \bibnamefont{and} \bibinfo{author}{\bibfnamefont{N.}~\bibnamefont{Menon}},
  \bibinfo{journal}{Science} \textbf{\bibinfo{volume}{317}},
  \bibinfo{pages}{105} (\bibinfo{year}{2007}).

\bibitem[{\citenamefont{Yang et~al.}(2010)\citenamefont{Yang, Marceau, and
  Gompper}}]{Gompper}
\bibinfo{author}{\bibfnamefont{Y.}~\bibnamefont{Yang}},
  \bibinfo{author}{\bibfnamefont{V.}~\bibnamefont{Marceau}}, \bibnamefont{and}
  \bibinfo{author}{\bibfnamefont{G.}~\bibnamefont{Gompper}},
  \bibinfo{journal}{Phys. Rev. E} \textbf{\bibinfo{volume}{82}},
  \bibinfo{pages}{031904} (\bibinfo{year}{2010}).

\bibitem[{\citenamefont{Chen et~al.}(2012)\citenamefont{Chen, Dong, Be'er,
  Swinney, and Zhang}}]{2012SwinEtAl}
\bibinfo{author}{\bibfnamefont{X.}~\bibnamefont{Chen}},
  \bibinfo{author}{\bibfnamefont{X.}~\bibnamefont{Dong}},
  \bibinfo{author}{\bibfnamefont{A.}~\bibnamefont{Be'er}},
  \bibinfo{author}{\bibfnamefont{H.~L.} \bibnamefont{Swinney}},
  \bibnamefont{and} \bibinfo{author}{\bibfnamefont{H.~P.} \bibnamefont{Zhang}},
  \bibinfo{journal}{Phys. Rev. Lett.} \textbf{\bibinfo{volume}{108}},
  \bibinfo{pages}{148101} (\bibinfo{year}{2012}).

\bibitem[{\citenamefont{Ginelli et~al.}(2010)\citenamefont{Ginelli, Peruani,
  B\"ar, and Chat\'e}}]{Ginelli_PRL10}
\bibinfo{author}{\bibfnamefont{F.}~\bibnamefont{Ginelli}},
  \bibinfo{author}{\bibfnamefont{F.}~\bibnamefont{Peruani}},
  \bibinfo{author}{\bibfnamefont{M.}~\bibnamefont{B\"ar}}, \bibnamefont{and}
  \bibinfo{author}{\bibfnamefont{H.}~\bibnamefont{Chat\'e}},
  \bibinfo{journal}{Phys. Rev. Lett.} \textbf{\bibinfo{volume}{104}},
  \bibinfo{pages}{184502} (\bibinfo{year}{2010}).

\bibitem[{\citenamefont{Sokolov and Aranson}(2012)}]{Sokolov_PRL12}
\bibinfo{author}{\bibfnamefont{A.}~\bibnamefont{Sokolov}} \bibnamefont{and}
  \bibinfo{author}{\bibfnamefont{I.~S.} \bibnamefont{Aranson}},
  \bibinfo{journal}{Phys. Rev. Lett.} \textbf{\bibinfo{volume}{109}},
  \bibinfo{pages}{248109} (\bibinfo{year}{2012}).

\bibitem[{\citenamefont{Saintillan and Shelley}(2008)}]{2008Saint_Shelley}
\bibinfo{author}{\bibfnamefont{D.}~\bibnamefont{Saintillan}} \bibnamefont{and}
  \bibinfo{author}{\bibfnamefont{M.~J.} \bibnamefont{Shelley}},
  \bibinfo{journal}{Phys. Fluids} \textbf{\bibinfo{volume}{20}},
  \bibinfo{pages}{123304} (\bibinfo{year}{2008}).

\bibitem[{\citenamefont{Wensink~et al}(2012)}]{Wensink_PNAS}
\bibinfo{author}{\bibfnamefont{H.~H.} \bibnamefont{Wensink~et al}},
  \bibinfo{journal}{Proc. Natl. Acad. Sci. USA} \textbf{\bibinfo{volume}{109}},
  \bibinfo{pages}{14308} (\bibinfo{year}{2012}).

\bibitem[{\citenamefont{Liu and I}(2013)}]{Li_PRE}
\bibinfo{author}{\bibfnamefont{K.-A.} \bibnamefont{Liu}} \bibnamefont{and}
  \bibinfo{author}{\bibfnamefont{L.}~\bibnamefont{I}}, \bibinfo{journal}{Phys.
  Rev. E} \textbf{\bibinfo{volume}{88}}, \bibinfo{pages}{033004}
  (\bibinfo{year}{2013}).

\bibitem[{\citenamefont{Yang et~al.}(2014)\citenamefont{Yang, Qiu, and
  Gompper}}]{yang2014}
\bibinfo{author}{\bibfnamefont{Y.}~\bibnamefont{Yang}},
  \bibinfo{author}{\bibfnamefont{F.}~\bibnamefont{Qiu}}, \bibnamefont{and}
  \bibinfo{author}{\bibfnamefont{G.}~\bibnamefont{Gompper}},
  \bibinfo{journal}{Phys. Rev. E} \textbf{\bibinfo{volume}{89}},
  \bibinfo{pages}{012720} (\bibinfo{year}{2014}).

\bibitem[{\citenamefont{Zhou et~al.}(2014)\citenamefont{Zhou, Sokolov,
  Lavrentovich, and Aranson}}]{zhou2014}
\bibinfo{author}{\bibfnamefont{S.}~\bibnamefont{Zhou}},
  \bibinfo{author}{\bibfnamefont{A.}~\bibnamefont{Sokolov}},
  \bibinfo{author}{\bibfnamefont{O.~D.} \bibnamefont{Lavrentovich}},
  \bibnamefont{and} \bibinfo{author}{\bibfnamefont{I.~S.}
  \bibnamefont{Aranson}}, \bibinfo{journal}{PNAS}
  \textbf{\bibinfo{volume}{111}}, \bibinfo{pages}{1265} (\bibinfo{year}{2014}).

\bibitem[{\citenamefont{Galajda et~al.}(2007)\citenamefont{Galajda, Keymer,
  Chaikin, and Austin}}]{chaikin}
\bibinfo{author}{\bibfnamefont{P.}~\bibnamefont{Galajda}},
  \bibinfo{author}{\bibfnamefont{J.}~\bibnamefont{Keymer}},
  \bibinfo{author}{\bibfnamefont{P.}~\bibnamefont{Chaikin}}, \bibnamefont{and}
  \bibinfo{author}{\bibfnamefont{R.}~\bibnamefont{Austin}},
  \bibinfo{journal}{J. Bacteriol.} \textbf{\bibinfo{volume}{189}},
  \bibinfo{pages}{8704} (\bibinfo{year}{2007}).

\bibitem[{\citenamefont{Wan et~al.}(2008)\citenamefont{Wan, Olson~Reichhardt,
  Nussinov, and Reichhardt}}]{ReichhardtPRL}
\bibinfo{author}{\bibfnamefont{M.~B.} \bibnamefont{Wan}},
  \bibinfo{author}{\bibfnamefont{C.~J.} \bibnamefont{Olson~Reichhardt}},
  \bibinfo{author}{\bibfnamefont{Z.}~\bibnamefont{Nussinov}}, \bibnamefont{and}
  \bibinfo{author}{\bibfnamefont{C.}~\bibnamefont{Reichhardt}},
  \bibinfo{journal}{Phys. Rev. Lett.} \textbf{\bibinfo{volume}{101}},
  \bibinfo{pages}{018102} (\bibinfo{year}{2008}).

\bibitem[{\citenamefont{Mi\~no~et al}(2011)}]{Clement_PRL11}
\bibinfo{author}{\bibfnamefont{G.}~\bibnamefont{Mi\~no~et al}},
  \bibinfo{journal}{Phys. Rev. Lett.} \textbf{\bibinfo{volume}{106}},
  \bibinfo{pages}{048102} (\bibinfo{year}{2011}).

\bibitem[{\citenamefont{Angelani et~al.}(2009)\citenamefont{Angelani,
  DiLeonardo, and Ruocco}}]{DiLeonardoPRL}
\bibinfo{author}{\bibfnamefont{L.}~\bibnamefont{Angelani}},
  \bibinfo{author}{\bibfnamefont{R.}~\bibnamefont{DiLeonardo}},
  \bibnamefont{and} \bibinfo{author}{\bibfnamefont{G.}~\bibnamefont{Ruocco}},
  \bibinfo{journal}{Phys. Rev. Lett.} \textbf{\bibinfo{volume}{102}},
  \bibinfo{pages}{048104} (\bibinfo{year}{2009}).

\bibitem[{\citenamefont{Angelani and DiLeonardo}(2010)}]{AngelaniCARGO}
\bibinfo{author}{\bibfnamefont{L.}~\bibnamefont{Angelani}} \bibnamefont{and}
  \bibinfo{author}{\bibfnamefont{R.}~\bibnamefont{DiLeonardo}},
  \bibinfo{journal}{New J. Phys.} \textbf{\bibinfo{volume}{12}},
  \bibinfo{pages}{113017} (\bibinfo{year}{2010}).

\bibitem[{\citenamefont{Sokolov et~al.}(2010)\citenamefont{Sokolov, Apodaca,
  Grzyboski, and Aranson}}]{SokolovPNAS}
\bibinfo{author}{\bibfnamefont{A.}~\bibnamefont{Sokolov}},
  \bibinfo{author}{\bibfnamefont{M.~M.} \bibnamefont{Apodaca}},
  \bibinfo{author}{\bibfnamefont{B.~A.} \bibnamefont{Grzyboski}},
  \bibnamefont{and} \bibinfo{author}{\bibfnamefont{I.~S.}
  \bibnamefont{Aranson}}, \bibinfo{journal}{Proc. Natl. Acad. Sci. USA}
  \textbf{\bibinfo{volume}{107}}, \bibinfo{pages}{969} (\bibinfo{year}{2010}).

\bibitem[{\citenamefont{Wong et~al.}(2013)\citenamefont{Wong, Beattie, Steager,
  and Kumar}}]{Wong}
\bibinfo{author}{\bibfnamefont{D.}~\bibnamefont{Wong}},
  \bibinfo{author}{\bibfnamefont{E.~E.} \bibnamefont{Beattie}},
  \bibinfo{author}{\bibfnamefont{E.~B.} \bibnamefont{Steager}},
  \bibnamefont{and} \bibinfo{author}{\bibfnamefont{V.}~\bibnamefont{Kumar}},
  \bibinfo{journal}{Appl. Phys. Lett.} \textbf{\bibinfo{volume}{103}},
  \bibinfo{eid}{153707} (\bibinfo{year}{2013}).

\bibitem[{\citenamefont{Wensink et~al.}(2014)\citenamefont{Wensink, Kantsler,
  Goldstein, and Dunkel}}]{wensink2014}
\bibinfo{author}{\bibfnamefont{H.~H.} \bibnamefont{Wensink}},
  \bibinfo{author}{\bibfnamefont{V.}~\bibnamefont{Kantsler}},
  \bibinfo{author}{\bibfnamefont{R.~E.} \bibnamefont{Goldstein}},
  \bibnamefont{and} \bibinfo{author}{\bibfnamefont{J.}~\bibnamefont{Dunkel}},
  \bibinfo{journal}{Phys. Rev. E} \textbf{\bibinfo{volume}{89}},
  \bibinfo{pages}{010302} (\bibinfo{year}{2014}).

\bibitem[{\citenamefont{DiLeonardo~et al}(2010)}]{di2010bacterial}
\bibinfo{author}{\bibfnamefont{R.}~\bibnamefont{DiLeonardo~et al}},
  \bibinfo{journal}{PNAS} \textbf{\bibinfo{volume}{107}}, \bibinfo{pages}{9541}
  (\bibinfo{year}{2010}).

\bibitem[{\citenamefont{Lauga et~al.}(2006)\citenamefont{Lauga, DiLuzio,
  Whitesides, and Stone}}]{lauga2006swimming}
\bibinfo{author}{\bibfnamefont{E.}~\bibnamefont{Lauga}},
  \bibinfo{author}{\bibfnamefont{W.~R.} \bibnamefont{DiLuzio}},
  \bibinfo{author}{\bibfnamefont{G.~M.} \bibnamefont{Whitesides}},
  \bibnamefont{and} \bibinfo{author}{\bibfnamefont{H.~A.} \bibnamefont{Stone}},
  \bibinfo{journal}{Biophys. Jour.} \textbf{\bibinfo{volume}{90}},
  \bibinfo{pages}{400} (\bibinfo{year}{2006}).

\bibitem[{\citenamefont{Li and Tang}(2009)}]{li2009accumulation}
\bibinfo{author}{\bibfnamefont{G.}~\bibnamefont{Li}} \bibnamefont{and}
  \bibinfo{author}{\bibfnamefont{J.~X.} \bibnamefont{Tang}},
  \bibinfo{journal}{Phys. Rev. Lett.} \textbf{\bibinfo{volume}{103}},
  \bibinfo{pages}{078101} (\bibinfo{year}{2009}).

\bibitem[{\citenamefont{Drescher et~al.}(2011)\citenamefont{Drescher, Dunkel,
  Cisneros, Ganguly, and Goldstein}}]{drescher2011fluid}
\bibinfo{author}{\bibfnamefont{K.}~\bibnamefont{Drescher}},
  \bibinfo{author}{\bibfnamefont{J.}~\bibnamefont{Dunkel}},
  \bibinfo{author}{\bibfnamefont{L.~H.} \bibnamefont{Cisneros}},
  \bibinfo{author}{\bibfnamefont{S.}~\bibnamefont{Ganguly}}, \bibnamefont{and}
  \bibinfo{author}{\bibfnamefont{R.~E.} \bibnamefont{Goldstein}},
  \bibinfo{journal}{Proc. Natl. Acad. Sci.} \textbf{\bibinfo{volume}{108}},
  \bibinfo{pages}{10940} (\bibinfo{year}{2011}).

\bibitem[{\citenamefont{Ishikawa~et al}(2011)}]{ishikawa2011energy}
\bibinfo{author}{\bibfnamefont{T.}~\bibnamefont{Ishikawa~et al}},
  \bibinfo{journal}{Phys. Rev. Lett.} \textbf{\bibinfo{volume}{107}},
  \bibinfo{pages}{028102} (\bibinfo{year}{2011}).

\bibitem[{\citenamefont{Calzavarini et~al.}(2008)\citenamefont{Calzavarini,
  Cencini, Lohse, and Toschi}}]{Lohse}
\bibinfo{author}{\bibfnamefont{E.}~\bibnamefont{Calzavarini}},
  \bibinfo{author}{\bibfnamefont{M.}~\bibnamefont{Cencini}},
  \bibinfo{author}{\bibfnamefont{D.}~\bibnamefont{Lohse}}, \bibnamefont{and}
  \bibinfo{author}{\bibfnamefont{F.}~\bibnamefont{Toschi}},
  \bibinfo{journal}{Phys. Rev. Lett.} \textbf{\bibinfo{volume}{101}},
  \bibinfo{pages}{084504} (\bibinfo{year}{2008}).

\bibitem[{\citenamefont{Gachelin~et a}(2013)}]{Clement_PRL13}
\bibinfo{author}{\bibfnamefont{J.}~\bibnamefont{Gachelin~et al}},
  \bibinfo{journal}{Phys. Rev. Lett.} \textbf{\bibinfo{volume}{110}},
  \bibinfo{pages}{268103} (\bibinfo{year}{2013}).

\bibitem[{\citenamefont{Sokolov et~al.}(2007)\citenamefont{Sokolov, Aranson,
  Kessler, and Goldstein}}]{Sokolov_PRL07}
\bibinfo{author}{\bibfnamefont{A.}~\bibnamefont{Sokolov}},
  \bibinfo{author}{\bibfnamefont{I.~S.} \bibnamefont{Aranson}},
  \bibinfo{author}{\bibfnamefont{J.~O.} \bibnamefont{Kessler}},
  \bibnamefont{and} \bibinfo{author}{\bibfnamefont{R.~E.}
  \bibnamefont{Goldstein}}, \bibinfo{journal}{Phys. Rev. Lett.}
  \textbf{\bibinfo{volume}{98}}, \bibinfo{pages}{158102}
  (\bibinfo{year}{2007}).

\bibitem[{\citenamefont{Sokolov et~al.}(2009)\citenamefont{Sokolov, Goldstein,
  Feldchtein, and Aranson}}]{Sokolov_PRE09}
\bibinfo{author}{\bibfnamefont{A.}~\bibnamefont{Sokolov}},
  \bibinfo{author}{\bibfnamefont{R.~E.} \bibnamefont{Goldstein}},
  \bibinfo{author}{\bibfnamefont{F.~I.} \bibnamefont{Feldchtein}},
  \bibnamefont{and} \bibinfo{author}{\bibfnamefont{I.~S.}
  \bibnamefont{Aranson}}, \bibinfo{journal}{Phys. Rev. E}
  \textbf{\bibinfo{volume}{80}}, \bibinfo{pages}{031903}
  (\bibinfo{year}{2009}).

\bibitem[{Sup()}]{SuppMat}
\eprint{See Supplemental Material at [URL] for movies and additional data.}

\bibitem[{\citenamefont{Kirchhoff et~al.}(1996)\citenamefont{Kirchhoff,
  L\"owen, and Klein}}]{Kirchhoff1996}
\bibinfo{author}{\bibfnamefont{T.}~\bibnamefont{Kirchhoff}},
  \bibinfo{author}{\bibfnamefont{H.}~\bibnamefont{L\"owen}}, \bibnamefont{and}
  \bibinfo{author}{\bibfnamefont{R.}~\bibnamefont{Klein}},
  \bibinfo{journal}{Phys. Rev. E} \textbf{\bibinfo{volume}{53}},
  \bibinfo{pages}{5011} (\bibinfo{year}{1996}).

\bibitem[{\citenamefont{Aranson et~al.}(2007)\citenamefont{Aranson, Sokolov,
  Kessler, and Goldstein}}]{aranson2007}
\bibinfo{author}{\bibfnamefont{I.~S.} \bibnamefont{Aranson}},
  \bibinfo{author}{\bibfnamefont{A.}~\bibnamefont{Sokolov}},
  \bibinfo{author}{\bibfnamefont{J.~O.} \bibnamefont{Kessler}},
  \bibnamefont{and} \bibinfo{author}{\bibfnamefont{R.~E.}
  \bibnamefont{Goldstein}}, \bibinfo{journal}{Phys. Rev. E}
  \textbf{\bibinfo{volume}{75}}, \bibinfo{pages}{040901}
  (\bibinfo{year}{2007}).

\bibitem[{\citenamefont{Dombrowski et~al.}(2004)\citenamefont{Dombrowski,
  Cisneros, Chatkaew, Goldstein, and Kessler}}]{dombrowski2004self}
\bibinfo{author}{\bibfnamefont{C.}~\bibnamefont{Dombrowski}},
  \bibinfo{author}{\bibfnamefont{L.}~\bibnamefont{Cisneros}},
  \bibinfo{author}{\bibfnamefont{S.}~\bibnamefont{Chatkaew}},
  \bibinfo{author}{\bibfnamefont{R.~E.} \bibnamefont{Goldstein}},
  \bibnamefont{and} \bibinfo{author}{\bibfnamefont{J.~O.}
  \bibnamefont{Kessler}}, \bibinfo{journal}{Phys. Rev. Lett.}
  \textbf{\bibinfo{volume}{93}}, \bibinfo{pages}{098103}
  (\bibinfo{year}{2004}).

\bibitem[{\citenamefont{Tirado et~al.}(1984)\citenamefont{Tirado, Martinez, and
  de~la Torre}}]{tirado}
\bibinfo{author}{\bibfnamefont{M.~M.} \bibnamefont{Tirado}},
  \bibinfo{author}{\bibfnamefont{C.~L.} \bibnamefont{Martinez}},
  \bibnamefont{and} \bibinfo{author}{\bibfnamefont{J.~G.} \bibnamefont{de~la
  Torre}}, \bibinfo{journal}{J. Chem. Phys.} \textbf{\bibinfo{volume}{81}},
  \bibinfo{pages}{2047} (\bibinfo{year}{1984}).

\bibitem[{\citenamefont{Garcia~de~la Torre
  et~al.}(1994)\citenamefont{Garcia~de~la Torre, Navarro, Lopez~Martinez, Diaz,
  and Lopez~Cascales}}]{delaTorreNMDC1994}
\bibinfo{author}{\bibfnamefont{J.}~\bibnamefont{Garcia~de~la Torre}},
  \bibinfo{author}{\bibfnamefont{S.}~\bibnamefont{Navarro}},
  \bibinfo{author}{\bibfnamefont{M.~C.} \bibnamefont{Lopez~Martinez}},
  \bibinfo{author}{\bibfnamefont{F.~G.} \bibnamefont{Diaz}}, \bibnamefont{and}
  \bibinfo{author}{\bibfnamefont{J.~J.} \bibnamefont{Lopez~Cascales}},
  \bibinfo{journal}{Biophys. J.} \textbf{\bibinfo{volume}{67}},
  \bibinfo{pages}{530} (\bibinfo{year}{1994}).

\bibitem[{\citenamefont{Carrasco and Garcia de~la Torre}(1999)}]{Carrasco99}
\bibinfo{author}{\bibfnamefont{B.}~\bibnamefont{Carrasco}} \bibnamefont{and}
  \bibinfo{author}{\bibfnamefont{J.}~\bibnamefont{Garcia de~la Torre}},
  \bibinfo{journal}{J. Chem. Phys.} \textbf{\bibinfo{volume}{111}},
  \bibinfo{pages}{4817} (\bibinfo{year}{1999}).

\bibitem[{\citenamefont{Wensink and L\"owen}(2012)}]{WensinkJPCM}
\bibinfo{author}{\bibfnamefont{H.~H.} \bibnamefont{Wensink}} \bibnamefont{and}
  \bibinfo{author}{\bibfnamefont{H.}~\bibnamefont{L\"owen}},
  \bibinfo{journal}{J. Phys.: Condens. Matter} \textbf{\bibinfo{volume}{24}},
  \bibinfo{pages}{464130} (\bibinfo{year}{2012}).

\bibitem[{\citenamefont{Sokolov and Aranson}(2009)}]{Sokolov_PRL09}
\bibinfo{author}{\bibfnamefont{A.}~\bibnamefont{Sokolov}} \bibnamefont{and}
  \bibinfo{author}{\bibfnamefont{I.~S.} \bibnamefont{Aranson}},
  \bibinfo{journal}{Phys. Rev. Lett.} \textbf{\bibinfo{volume}{103}},
  \bibinfo{pages}{148101} (\bibinfo{year}{2009}).

\end{thebibliography}
\end{document}